\documentclass{llncs}
\usepackage{a4wide}
\usepackage{amsmath}
\usepackage{amssymb}
\usepackage{bbold}
\usepackage{mathtools}
\usepackage{stackengine}
\usepackage{xcolor,colortbl}
\definecolor{Yellow}{rgb}{1,1,0}
\newcolumntype{y}{>{\columncolor{Yellow}}c}
\stackMath
\title{Baugh-Wooley Multiplication for the RISCV Processor}
\author{F.A. Grootjen\orcidID{0000-0002-0548-5831} \and N.K. Schauer\orcidID{0009-0000-2211-8083}}
\institute{Artificial Intelligence, Radboud University, Nijmegen, the Netherlands}
\begin{document}
\maketitle
\abstract{This article describes an efficient way to implement the multiplication instructions for a RISCV processor. Instead
of using three predefined IP blocks for signed, unsigned and mixed multiplication, this article presents a novel extension to the Baugh-Wooley
multiplication algorithm which reduces area and power consumption with roughly a factor three.
}
\section{Introduction}
Within a RISC instruction set, the multiplication instruction is a bit of an oddball. While almost all instructions
use only one clock cycle, the multiplication instruction uses significant resources which is notable
when looking at the implementation's latency and/or density. Within the RISCV instruction set architecture, the multiplication instruction is not
even present in the \emph{basic} instruction set, allowing low end implementations 
to adhere to the basic specs \cite{Waterman16}. Of course, most (larger)
implementations support the M extension which do provide multiplication (and division) instructions.
Within this M extension, there are unsigned, signed and mixed multiplication variants. At first, it seems to be strange
to have these different versions: for example when looking at the 32 bit multiplication instructions, the signedness 
of the operands does not matter for the result\footnote{In this article we will show that this is
the case. Suprisingly it is not as straightforward as you would expect.}. But this only is true for
the \emph{lower 32 bits} of the result, for the higher 32 bits
the signedness of the operands does matter.

Most modern multiplication implementations use some adapted form of \emph{long multiplication} (see \ref{longmultiplication}).
These implementations have two distinct phases:

\begin{enumerate}
  \item Calculation of partial products
  \item Summation of all partial products
\end{enumerate}

While first phase is pretty straightforward and easily parallelized, the second phase is more complex. Literature contains
many solutions for the second phase that cover the traditional speed/area tradoff: from simple binary adder trees to more sophisticated 
\cite{Sakthikumaran11,Swathi21} and
area expensive solutions
\cite{Wallace64,Dadda65}.

The goal of this paper is to show how small adaptations to phase 1 make it possible to share phase 2 for all three multiplication forms: signed, unsigned
and mixed. Section~\ref{unsigned} introduces long multiplication and shows how it works on binary unsigned sequences. Section~\ref{signed} derives Baugh and Wooleys adaptation to use signed numbers. Section
\ref{mixed} explains how a similar adaptation can be done for the mixed signed numbers. Finally Section~\ref{finally} wraps them all together
and shows that a single multiplicator can be used for all RV32 multiplication instructions.
\section{Multiplication}\label{unsigned}
The mathematical definition of multiplication is based on \emph{repeated addition}:
\begin{definition}[Multiplication]
For $a\in\mathbb{N}$ and $b\in\mathbb{Z}$ the product $a\cdot b$ is defined as:
\begin{align*}
a\cdot b = \underbrace{b+b+\cdots +b}_{\textrm{$a$ times}}=\sum_{i=1}^{a}b
\end{align*}
\end{definition}
In this definition, the number $a$ is called the \emph{multiplier} and $b$ is called \emph{multiplicand}. The result $a \cdot b$ is
called \emph{product}. Note that for $a=0$ the product is $0$ by definition. Furthermore we can extend this definition for $a\in\mathbb{Z}$
 by defining: $a\cdot b=-((-a)\cdot b)$ for $a<0$.

On a computer, integers are limited in size. In the RV32 instruction set integers are 32 bits in size. So, multiplication takes two 32 bits integers
and produces a 64 bit product. The lower 32 bits of the product is stored by the \texttt{mul} instruction. For the higher 32 bits, the
specific instruction depends on the number format (signed or unsigned or mixed).

Integers are stored as binary bit sequences. For unsigned numbers the following format is used:
\begin{definition}[Unsigned numbers]
The unsigned number $a\in\mathbb{N}$ is represented as a binary sequence $a_{n-1}a_{n-2}\ldots a_1a_0$ with:
\begin{align*}
a = \sum_{i=0}^{n-1} a_i\cdot 2^i
\end{align*}
\end{definition}
Where $a_i\in\{0,1\}$ and $n$ denotes the number bits. Note that the largest number that can be represented this way is $2^n-1$. For 32 bit
numbers this is 4294967295.
\subsection{Long Multiplication}\label{longmultiplication}
Unsigned multiplication of binary sequences can be done using \emph{long multiplication}, but
now in a binary fashion. Table \ref{long4bit} shows the long multiplication of two 4 bit numbers: $1001$ (9) and $0101$ (5).
\begin{table}[htpb]
\begin{center}
\begin{tabular}{cccccccccl}
  &   &   &   &   & 1 & 0 & 0 & 1 & (=9)\\ 
  &   &   &   &$\times$ & 0 & 1 & 0 & 1 & (=5)\\ \hline
  &   &   &   &   & 1 & 0 & 0 & 1 &\\
  &   &   &   & 0 & 0 & 0 & 0 &   &\\ 
  &   &   & 1 & 0 & 0 & 1 &   &   &\\
+ &   & 0 & 0 & 0 & 0 &   &   &   &\\ \hline
  & 0 & 0 & 1 & 0 & 1 & 1 & 0 & 1 & (=45)\\
\end{tabular}
\end{center}
\caption{Long multiplication of two 4 bit numbers}\label{long4bit}
\end{table}

The long multiplication algorithm for binary sequences is straightforward: starting from the least significant bit of
the multiplier write down the multiplicant only when the bit is 1, otherwise write down zeroes. We repeat this procedure for the other
bits (at position $i$), and write down the multiplicant shifted $i$ bits to the left. This way we get $n$ partial products.
The final product is the sum of all partial products. It is relatively easy to backup the long multiplication mathematically:
\begin{align*}
a\cdot b=\left(\sum_{i=0}^{n-1} a_i\cdot 2^i \right)\cdot b=b\cdot \sum_{i=0}^{n-1} a_i\cdot 2^i=
\sum_{i=0}^{n-1} b\cdot a_i\cdot 2^i
\end{align*}
As you can see the product is a sum of $n$ partial products. Within the partial product $2^i$ represents the left shift and since $a_i$ is either 0 or 1, the
product $b\cdot a_i$ is either all zeroes or the multiplicand itself.
\subsection{Hardware implementation}
Performing the above algorithm in hardware involves 2 tasks:

\begin{enumerate}
  \item calculate the $n$ partial products, which can be calculated in parallel
  \item calculate the sum of all partial products, using for example an adder tree
\end{enumerate}

For a partial product we have to calculate $b\cdot a_i \cdot 2^i$. As stated above, $2^i$ is actually a \emph{fixed} shift (depending on $i$).
Since $a_i \in \{0,1\}$ the multiplication with $b$ can be done with an \texttt{and} vector operation after sign-extending $a_i$ to a vector of length $n$.
\subsection{Example}
For 4x4 to 8 bit, the unsigned multiplication scheme looks as follows:
\begin{center}
\begin{tabular}{cccccccccl}
  &   &   &   &         & $b_3$ & $b_2$ & $b_1$ & $b_0$\\ 
  &   &   &   &$\times$ & $a_3$ & $a_2$ & $a_1$ & $a_0$ \\ \hline
  &   &   &   &   & $a_0b_3$ & $a_0b_2$ & $a_0b_1$ & $a_0b_0$\\
  &   &   &   & $a_1b_3$ & $a_1b_2$ & $a_1b_1$ & $a_1b_0$ &  \\ 
  &   &   & $a_2b_3$ & $a_2b_2$ & $a_2b_1$ & $a_2b_0$ &   &  \\
+ &   & $a_3b_3$ & $a_3b_2$ & $a_3b_1$ & $a_3b_0$ &   &   &  \\ \hline
  &  &  &  & \ldots &   &  &  & \\
\end{tabular}
\end{center}
\section{Signed Multiplication}\label{signed}
Signed numbers are stored as binary sequences with a twist:
\begin{definition}[Signed numbers]
The (2's complement) signed number $a\in\mathbb{Z}$ is represented as a binary sequence $a_{n-1}a_{n-2}\ldots a_1a_0$ with:
\begin{align*}
a = -a_0\cdot 2^{n-1}+\sum_{i=0}^{n-2} a_i\cdot 2^i
\end{align*}
\end{definition}
Where $a_i\in\{0,1\}$ and $n$ denotes the number of bits. Note that the smallest negative (largest positive) number that can be
represented this way is $-2^{n-1}$ ($2^{n-1}-1$ respectively). For 32 bit
numbers this is \mbox{-2147483648} (2147483647 respectively).

This way of denoting 2's complement signed numbers has the advantage that normal addition (and subtraction) still functions correctly. For signed
multiplication however it does not.
Baugh and Wooley \cite{Baugh73} found a nice way to use unsigned multiplication for signed numbers with only a small
number of changes. Let us derive their solution:
\begin{align*}
a\cdot b=\left(-a_{n-1}\cdot 2^{n-1}+\sum_{i=0}^{n-2} a_i\cdot 2^i\right) \cdot
         \left(-b_{n-1}\cdot 2^{n-1}+\sum_{j=0}^{n-2} b_i\cdot 2^j\right) =
\end{align*}
\begin{align*}
\underbrace{\vphantom{\sum_{j=0}^{n-2}}a_{n-1}b_{n-1}\cdot 2^{2n-2}}_{A}+\underbrace{\sum_{i=0}^{n-2}\sum_{j=0}^{n-2} a_ib_j\cdot 2^{i+j}}_{B}
 -\underbrace{\vphantom{\sum_{j=0}^{n-2}}2^{n-1}\sum_{i=0}^{n-2}a_ib_{n-1}\cdot 2^i}_{X}
 -\underbrace{2^{n-1}\sum_{j=0}^{n-2}a_{n-1}b_j\cdot 2^j}_{Y}
\end{align*}
Note that the first two terms ($A$ and $B$) are positive, so they do not pose a problem.
The last two terms are negative, so we are finding their 2's complement counterparts so we can simply add them.

Call the first negative term (without the sign) $X$ and the second one $Y$. We will focus on $X$ for now, for $Y$ the derivation is similar. So:
\begin{align*}
X=2^{n-1}\sum_{i=0}^{n-2}a_ib_{n-1}\cdot 2^i
\end{align*}
Since we are interested in the binary notation of $X$ define:
\begin{align*}
x_i=a_ib_{n-1}
\end{align*} 
Assuming $X$ to be $2n$ bits wide, we can write out its binary notation:
\begin{center}
\begin{tabular}{l|ccccccccccc}
bit position  & $2n-1$ & $2n-2$ & $2n-3$    & $2n-4$    & \ldots & $n$   & $n-1$   & $n-2$ & $n-3$ & \ldots & 0 \\ \hline
bit value $X$ & 0      & 0      & $x_{n-2}$ & $x_{n-3}$ & \ldots & $x_1$ & $x_0$   & 0     & 0     & \ldots & 0 \\ 
\end{tabular}
\end{center}
The 2's complement of $X$ can be calculated by inversing all bits and adding 1:
\begin{center}
\begin{tabular}{l|ccccccccccc}
bit position & $2n-1$ & $2n-2$ & $2n-3$    & $2n-4$    & \ldots & $n$   & $n-1$   & $n-2$ & $n-3$ & \ldots & 0 \\ \hline
bit value $-X$   & 1      & 1      & $\overline{x}_{n-2}$ & $\overline{x}_{n-3}$ & \ldots & $\overline{x}_1$ & $\overline{x}_0$   & 1     & 1     & \ldots & 1+1 \\ 
\end{tabular}
\end{center}
Note that the bit value row is not final, the addition at bitposition 0 will generate a carry that will cascade all the way
up to position $n-1$:
\begin{center}
\begin{tabular}{l|ccccccccccc}
bit position & $2n-1$ & $2n-2$ & $2n-3$    & $2n-4$    & \ldots & $n$   & $n-1$   & $n-2$ & $n-3$ & \ldots & 0 \\ \hline
bit value $-X$   & 1      & 1      & $\overline{x}_{n-2}$ & $\overline{x}_{n-3}$ & \ldots & $\overline{x}_1$ & $\overline{x}_0+1$   & 0     & 0     & \ldots & 0 \\ 
\end{tabular}
\end{center}
Similarly, we will find for $-Y$:
\begin{center}
\begin{tabular}{l|ccccccccccc}
bit position & $2n-1$ & $2n-2$ & $2n-3$    & $2n-4$    & \ldots & $n$   & $n-1$   & $n-2$ & $n-3$ & \ldots & 0 \\ \hline
bit value $-Y$  & 1      & 1      & $\overline{y}_{n-2}$ & $\overline{y}_{n-3}$ & \ldots & $\overline{y}_1$ & $\overline{y}_0+1$   & 0     & 0     & \ldots & 0 \\ 
\end{tabular}
\end{center}
When adding $-X$ and $-Y$ some bit positions are worth noticing:
\begin{center}
\begin{tabular}{l|cc|cc}
bit position & $2n-1$ & $2n-2$ & $n$   & $n-1$ \\ \hline
bit value $-X$   & 1      & 1      & $\overline{x}_1$ & $\overline{x}_0+1$ \\ 
bit value $-Y$   & 1      & 1      & $\overline{y}_1$ & $\overline{y}_0+1$ \\ 
bit value $-X-Y$ & 1      & 0      & $\overline{x}_1+\overline{y}_1+1$ & $\overline{x}_0+\overline{y}_0$ \\ 
\end{tabular}
\end{center}
So adding $-X-Y$ can be done by adding the inversed $x$ and $y$ bits, and subsequently adding
the bits with bit positions $2n-1$ and $n$.
\subsection{Example}
For 4x4 to 8 bit, the signed multiplication scheme looks as follows:
\begin{center}
\begin{tabular}{cccccccccc}
  &   &   &   &         & $b_3$ & $b_2$ & $b_1$ & $b_0$ \\ 
  &   &   &   &$\times$ & $a_3$ & $a_2$ & $a_1$ & $a_0$ \\ \hline
  &   & $a_3b_3$  &       &           &          &          &          &         & $A$ \\
  &   &           &       &           &          & $a_0b_2$ & $a_0b_1$ & $a_0b_0$& $B$ \\ 
  &   &           &       &           & $a_1b_2$ & $a_1b_1$ & $a_1b_0$ &         & $B$ \\ 
  &   &           &       & $a_2b_2$  & $a_2b_1$ & $a_2b_0$ &          &         & $B$ \\ 
  &   &           & $\overline{a_3b_2}$ & $\overline{a_3b_1}$ & $\overline{a_3b_0}$ & & & & $Y$ \\ 
  &   &           & $\overline{a_2b_3}$ & $\overline{a_1b_3}$ & $\overline{a_0b_3}$ & & & & $X$ \\ 
+ & 1  &  &  & $1$ &  &   &   &  & $X,Y$ \\ \hline
  &  &  &  & \ldots &   &  &  & \\
\end{tabular}
\end{center}
Squeezing things together gives:
\begin{center}
\begin{tabular}{ccccccccc}
  &   &   &   &         & $b_3$ & $b_2$ & $b_1$ & $b_0$ \\ 
  &   &   &   &$\times$ & $a_3$ & $a_2$ & $a_1$ & $a_0$ \\ \hline \\[-5pt]
  &   &           &       & $1$ & $\overline{a_0b_3}$ & $a_0b_2$ & $a_0b_1$ & $a_0b_0$\\ 
  &   &           &       & $\overline{a_1b_3}$ & $a_1b_2$ & $a_1b_1$ & $a_1b_0$ &         \\ 
  &   &           &  $\overline{a_2b_3}$ & $a_2b_2$  & $a_2b_1$ & $a_2b_0$ &          &         \\ 
+ & 1  & $a_3b_3$  & $\overline{a_3b_2}$ & $\overline{a_3b_1}$ & $\overline{a_3b_0}$ & & & \\ \hline
  &  &  &  & \ldots &   &  &  & \\
\end{tabular}
\end{center}
While comparing this scheme with the unsigned one (see Section~\ref{unsigned}) it is easy to see the similarities. If we only consider the lower
$n$ bits of the output, both schemes produce the same output. At first this is not obvious, since the unsigned scheme uses
$a_0b_3$ and $a_3b_0$ while the signed scheme has $\overline{a_0b_3}$ and $\overline{a_3b_0}$. Still the sum
(not the carry) is the same, see the following table:
\begin{center}
\begin{tabular}{cccycccy}
$a_0b_3$ & $a_3b_0$ & carry & sum & $\overline{a_0b_3}$ & $\overline{a_3b_0}$ & carry & sum \\ \hline
0 & 0 & 0 & 0 & 1 & 1 & 1 & 0 \\
0 & 1 & 0 & 1 & 1 & 0 & 0 & 1 \\
1 & 0 & 0 & 1 & 0 & 1 & 0 & 1 \\
1 & 1 & 1 & 0 & 0 & 0 & 0 & 0 \\
\end{tabular}
\end{center}
\section{Mixed Multiplication}\label{mixed}
The RISCV instruction set \cite{Waterman16} has a special instruction which takes an \emph{unsigned} multiplier and a
\emph{signed} multiplicant:
\begin{verbatim}
MULHSU rd, rs1, rs2
\end{verbatim}
which multiplies signed operand \texttt{rs1} (multiplicand) and unsigned operand \texttt{rs2} (multiplier) and stores
the upper 32 bits of the product in \texttt{rd}. We will try to derive an extension to Baugh-Wooley multiplication for mixed operands.
Let $a$ be the unsigned multiplier and $b$ the signed multiplicant.
\begin{align*}
a\cdot b=\left(\sum_{i=0}^{n-1} a_i\cdot 2^i\right) \cdot
         \left(-b_{n-1}\cdot 2^{n-1}+\sum_{j=0}^{n-2} b_i\cdot 2^j\right) =
\end{align*}
\begin{align*}
\underbrace{\sum_{i=0}^{n-1}\sum_{j=0}^{n-2} a_ib_j\cdot 2^{i+j}}_{B}
 -\underbrace{\vphantom{\sum_{j=0}^{n-2}}2^{n-1}\sum_{i=0}^{n-1}a_ib_{n-1}\cdot 2^i}_{X}
\end{align*}
The first term ($B$) is positive, so doesn't pose any problem. The second term $X$ is negative, so we are going to find its
2's complement notation so we can add it.
Since we are interested in the binary notation of $X$ define:
\begin{align*}
x_i=a_ib_{n-1}
\end{align*} 
Assuming $X$ to be $2n$ bits wide, we can write out its binary notation:
\begin{center}
\begin{tabular}{l|cccccccccc}
bit position  & $2n-1$ & $2n-2$ & $2n-3$    & \ldots & $n$   & $n-1$   & $n-2$ & $n-3$ & \ldots & 0 \\ \hline
bit value $X$ & 0      & $x_{n-1}$  & $x_{n-2}$ & \ldots & $x_1$ & $x_0$   & 0     & 0     & \ldots & 0 \\ 
\end{tabular}
\end{center}
The 2's complement of $X$ can be calculated by inversing all bits and adding 1:
\begin{center}
\begin{tabular}{l|ccccccccccc}
bit position & $2n-1$ & $2n-2$ & $2n-3$    & \ldots & $n$   & $n-1$   & $n-2$ & $n-3$ & \ldots & 0 \\ \hline
bit value $-X$   & 1      & $\overline{x}_{n-1}$ & $\overline{x}_{n-2}$ & \ldots & $\overline{x}_1$ & $\overline{x}_0$   & 1     & 1     & \ldots & 1+1 \\ 
\end{tabular}
\end{center}
Note that the bit value row is not final, the addition at bitposition 0 will generate a carry that will cascade all the way
up to position $n-1$:
\begin{center}
\begin{tabular}{l|ccccccccccc}
bit position & $2n-1$ & $2n-2$ & $2n-3$    & \ldots & $n$   & $n-1$   & $n-2$ & $n-3$ & \ldots & 0 \\ \hline
bit value $-X$   & 1      & $\overline{x}_{n-1}$ & $\overline{x}_{n-2}$ & \ldots & $\overline{x}_1$ & $\overline{x}_0+1$   & 0     & 0     & \ldots & 0 \\ 
\end{tabular}
\end{center}
\subsection{Example}
For 4x4 to 8 bit, the mixed multiplication scheme looks as follows:
\begin{center}
\begin{tabular}{cccccccccc}
  &   &   &   &         & $b_3$ & $b_2$ & $b_1$ & $b_0$ \\ 
  &   &   &   &$\times$ & $a_3$ & $a_2$ & $a_1$ & $a_0$ \\ \hline
  &   &           &       &           &          & $a_0b_2$ & $a_0b_1$ & $a_0b_0$& $B$ \\ 
  &   &           &       &           & $a_1b_2$ & $a_1b_1$ & $a_1b_0$ &         & $B$ \\ 
  &   &           &       & $a_2b_2$  & $a_2b_1$ & $a_2b_0$ &          &         & $B$ \\ 
  &   &           & $a_3b_2$ & $a_3b_1$ & $a_3b_0$ & & & & $B$ \\ 
  &   & $\overline{a_3b_3}$ & $\overline{a_2b_3}$ & $\overline{a_1b_3}$ & $\overline{a_0b_3}$ & & & & $X$ \\ 
+ & 1  &  &  &  & $1$ &   &   &  & $X$ \\ \hline
  &  &  &  & \ldots &   &  &  & \\
\end{tabular}
\end{center}
Squeezing things together gives:
\begin{center}
\begin{tabular}{ccccccccc}
  &   &   &   &         & $b_3$ & $b_2$ & $b_1$ & $b_0$ \\ 
  &   &   &   &$\times$ & $a_3$ & $a_2$ & $a_1$ & $a_0$ \\ \hline \\[-5pt]
  &   &           &       &     & $\overline{a_0b_3}$ & $a_0b_2$ & $a_0b_1$ & $a_0b_0$\\ 
  &   &           &       & $\overline{a_1b_3}$ & $a_1b_2$ & $a_1b_1$ & $a_1b_0$ &         \\ 
  &   &           &  $\overline{a_2b_3}$ & $a_2b_2$  & $a_2b_1$ & $a_2b_0$ &          &         \\ 
  & 1  & $\overline{a_3b_3}$  & $a_3b_2$ & $a_3b_1$ & $a_3b_0$ & & & \\ 
+ &    &  &  &  & $1$ &   &   &  \\ \hline
  &  &  &  & \ldots &   &  &  & \\
\end{tabular}
\end{center}
Having the last row (adding a 1 to bit position $n-1$) is a bit unfortunate. If we plan to perform all additions with a tree adder this single line
results in an extra (non parallel) action. However, there is a trick to squeeze the scheme a bit more:
\begin{center}
\begin{tabular}{ccccccccc}
  &   &   &   &         & $b_3$ & $b_2$ & $b_1$ & $b_0$ \\ 
  &   &   &   &$\times$ & $a_3$ & $a_2$ & $a_1$ & $a_0$ \\ \hline \\[-5pt]
  &   &           &       & $\overline{a_0b_3}$ & $a_0b_3$ & $a_0b_2$ & $a_0b_1$ & $a_0b_0$\\ 
  &   &           &       & $\overline{a_1b_3}$ & $a_1b_2$ & $a_1b_1$ & $a_1b_0$ &         \\ 
  &   &           &  $\overline{a_2b_3}$ & $a_2b_2$  & $a_2b_1$ & $a_2b_0$ &          &         \\ 
+ & 1  & $\overline{a_3b_3}$  & $a_3b_2$ & $a_3b_1$ & $a_3b_0$ & & & \\ \hline
  &  &  &  & \ldots &   &  &  & \\
\end{tabular}
\end{center}
The correctness of this last step is easy to see: for all values of $a_0b_3$ it is equal to $\overline{a_0b_3}+1$ (disregarding the carry).
The carry itself equals $\overline{a_0b_3}$.
\begin{center}
\begin{tabular}{|y|c|c|y|} \hline
         &                     & \multicolumn{2}{c|}{} \\
         &                     & \multicolumn{2}{c|}{$\overline{a_0b_3}+1$} \\
$a_0b_3$ & $\overline{a_0b_3}$ & carry & sum \\ \hline
0        & 1                   &   1   &  0  \\
1        & 0                   &   0   &  1  \\ \hline
\end{tabular}
\end{center}
\section{Merging}\label{finally}
Assuming we have binary signals $s, u$ and $m$ representing signed, unsigned and mixed multiplication respectively (possibly decoded 
with a demux from the RISCV multiplication instruction, see Table~\ref{multable}) we can now merge all implementations into a single
multiplier\footnote{Note that we do not actually use the $u$ signal but exploit the fact that for unsigned multiplication both $s$ as $m$ are 0.}. 

\begin{table}
\begin{center}
\begin{tabular}{|c|c|ccc|} \hline
instruction     & description & $s$ & $u$ & $m$ \\ \hline
\texttt{mulh}   & signed      & 1   & 0 & 0 \\
\texttt{mulhu}  & unsigned    & 0   & 1 & 0 \\
\texttt{mulhsu} & mixxed      & 0   & 0 & 1 \\ \hline
\end{tabular}
\end{center}
\caption{RV32 multiplication instructions and their demuxed signals}\label{multable}
\end{table}

Essentially there are 3 different types of partial products:
the first one, the last one and all in between. We will use $\oplus$ for the bitwise xor operator. Unmentioned bit positions are considered 0.
\subsection{The First Partial Product}
\begin{center}
\begin{tabular}{l|cc|cc|cc|cc|cc|cc}
result bit    &&  $s + m \overline{a_0b_{n-1}}$ &&  $s \oplus a_0b_{n-1}$ && $a_0b_{n-2}$ && \ldots && $a_0b_1$ && $a_0b_0$ \\ \hline
bit position  && $n$                                && $n-1$                      && $n-2$            && \ldots && 1        && 0 \\
\end{tabular}
\end{center}
As you can see, the first partial product is standard (unchanged) for bit positions 0 up to and including $n-2$. Bit position $n-1$ is inverted
for signed multiplications while
bit position $n$ is 1 for signed and the inverted value of $a_0b_{n-1}$ for mixed multiplications.
\subsection{The Intermediate Partial Products}
The intermediate partial products (from 1 upto $n-2$) are shifted depending on their (fixed) number.
\begin{center}
\begin{tabular}{l|cc|cc|cc|cc|cc}
result bit     && $(s+m) \oplus a_ib_{n-1}$  && $a_ib_{n-2}$ && \ldots && $a_ib_1$ && $a_ib_0$ \\ \hline
bit position   && $n-1+i$                    && $n-2+i$      && \ldots && i+1      && i \\
\end{tabular}
\end{center}
Apart from being shifted, the bit values are unchanged. Only the most signigicant bit is inverted when performing a signed or mixed
multiplication.
\subsection{The Last Partial Product}
\begin{center}
\begin{tabular}{l|cc|cc|cc|cc|cc|cc}
result bit     && $s+m$  && $m \oplus a_{n-1}b_{n-1}$ && $s\oplus a_{n-1}b_{n-2}$ && \ldots && $s\oplus a_{n-1}b_1$ && $s\oplus a_{n-1}b_0$ \\ \hline
bit position   && $2n-1$ && $2n-2$                    && $2n-3$                   && \ldots && n-2                 && n-1 \\
\end{tabular}
\end{center}
For the last partial product we have to invert its bitvalues when dealing with signed multiplication. The most significant bit however should only be
inverted for mixed multiplications. Finally, both in signed and mixed multiplications the answer should be preceded with a 1 (bit position $2n-1$).
\section{Conclusion}
We showed that with some small changes to the partial products, all RV32 multiplications can be handled by a single multiplier,
reducing the implementation's complexity roughly with a factor three. This might be specifically benificial in multi-core and vectorlike implementations.
\bibliographystyle{splncs04}
\bibliography{bibliography}
\end{document}